\documentclass[%
 reprint, 
 amsmath,assume,
 aps,
pra,
floatfix,
]{revtex4-1}
\usepackage[utf8]{inputenc}
\usepackage[english]{babel}
\usepackage[T1]{fontenc}
\usepackage{amsmath}
\usepackage{xcolor} 

\definecolor{sph}{rgb}{0.0588, 0.3216, 0.7294} 
\definecolor{ppk}{rgb}{1.0, 0.4549, 0.0902} 
\definecolor{darkgray}{rgb}{0.66, 0.66, 0.66}
\definecolor{darkbrown}{rgb}{0.4, 0.26, 0.13}

\usepackage[colorlinks=true, linkcolor=ppk, citecolor=blue, urlcolor=sph]{hyperref}

\usepackage{tikz}
\usepackage{lipsum}

\usepackage{graphicx}

\usepackage{ifthen}
\newcommand{\nc}{\newcommand}
\nc{\nn}{\nonumber}
\nc{\txt}{\textrm}
\nc{\txtsup}{\textsuperscript}
\nc{\txtsub}{\textsubscript}
\nc{\calL}{\mathcal{L}}
\nc{\U}{\mathcal{U}}
\nc{\T}{\mathcal{T}}
\nc{\E}{\mathcal{E}}
\nc{\calH}{\mathcal{H}}
\nc{\ave}[1][]{%
    \ifthenelse{\equal{#1}{}}%
    {\langle \rangle}
    {\langle #1 \rangle}
}

\begin{document}
\title{Protecting coherence from the environment via Stark many-body localization in a Quantum-Dot Simulator}

\author{Subhajit Sarkar}
\email{sbhjt72@gmail.com}
\email{subhajis@srmist.edu.in}
\affiliation{Department of Physics and Nanotechnology,
SRM Institute of Science and Technology
Kattankulathur-603 203, India}
\affiliation{Department of Chemistry and School of Electrical and Computer Engineering, Ben-Gurion University of the Negev, Beer-Sheva 84105, Israel}
\author{Berislav Bu\v{c}a}
\email{berislav.buca@physics.ox.ac.uk}
\affiliation{ Clarendon Laboratory, University of Oxford, Parks Road, Oxford OX1 3PU, United Kingdom}
\affiliation{Niels Bohr International Academy, Niels Bohr Institute, Copenhagen University, Universitetsparken 5, 2100 Copenhagen, Denmark}

\begin{abstract}
Semiconductor platforms are emerging as a promising architecture for storing and processing quantum information, e.g., in quantum dot spin qubits. However, charge noise coming from interactions between the electrons is a major limiting factor, along with the scalability of many qubits, for a quantum computer. We show that a magnetic field gradient can be implemented in a semiconductor quantum dot array to induce a local quantum coherent dynamical $\ell-$bit exhibiting the potential to be used as logical qubits. These dynamical $\ell-$bits are responsible for the model being many-body localized. We show that these dynamical $\ell-$bits and the corresponding many-body localization are protected from all noises, including phonons, for sufficiently long times if electron-phonon interaction is not non-local. We further show the implementation of thermalization-based self-correcting logical gates. This thermalization-based error correction goes beyond the standard paradigm of decoherence-free and noiseless subsystems. Our work thus opens a new venue for passive quantum error correction in semiconductor-based quantum computers. 
\end{abstract}
\maketitle

\section{Introduction}
Quantum coherence plays a significant role in physics and quantum information processing  \cite{DIVINCENZO1999202}, such as quantum cryptography \cite{quantum_crypto_RevModPhys,Pirandola:20, Grosshans2003, Coles2016}, metrology \cite{Giovannetti2011}, nanoscale thermodynamics \cite{Taniguchi_PhysRevB, Horodecki2013}, and in much-debated energy transport in biological systems \cite{Romero2014, Harush_dubi_Sci_Adv}. The stable operation of the fundamental building block of quantum information, the qubit, relies on quantum coherence. However, the interaction of the qubits with other qubits and their environment is usually detrimental to the long-lived coherence necessary for qubit operations \cite{Bader2014}. 

To this end, many-body localized (MBL) systems have been argued to be a viable option due to the slow growth of entanglement leading to a relatively long decoherence time-scale, where the existence of {\sl $\ell-$bits} (representing intensive conservation laws) has the potential to be used as logical qubits \cite{Yao_arxiv_2015, Nandkishore_annurev-conmatphys_2015, Abanin_MBL, lbits1, Serbyn, Vosk_PhysRevX, Agarwal_PhysRevLett, ZnidaricMBL2, Bar_Lev_PhysRevLett, DaleyMBL, MBLexp, ZnidaricMBL,MBLent2,MBLent3,MBLEnt4}. However, the most studied disorder-induced MBL $\ell-$bits exhibit dephasing \footnote{By dephasing here we mean that the numerous $\ell$-bits coming into the dynamics with incommensurate (random) frequencies destructively interfere with the noise dynamics rather than coherent dynamics.} due to random frequencies coming from the disorder, making them sub-optimal for hosting quantum coherent structures like qubits \cite{MBLdephasing1}.

Recently, disorder-free MBL, also known as Stark MBL (SMBL) has been demonstrated both theoretically and experimentally \cite{Stark1, Stark2, guo2020stark, devendra_SMBL1, devendra_SMBL2, Wang}. Such an MBL occurs due to an applied gradient field to an otherwise translationally invariant system and exhibits oscillations of various many-body observables in numerical calculations \cite{Stark2, Ribeiro}, and in ultra-cold atom and trapped ion experiments \cite{Scherg, morong2021observation}, along with Hilbert-space fragmentation \cite{Scherg, StarkFrag2} and quantum many-body scars \cite{PapicStark}.

SMBL too exhibits spatially quasi-localized dynamical $\ell-$bits that are stable over a long time, with decay time-scale that is exponentially long in the (Stark) field gradient \cite{Buca_l-bit}. Consequently, the SMBL system shows many-body Bloch oscillations in correlation functions at frequencies given by the dynamical {\sl $\ell-$bits} \cite{Buca_l-bit, Buca, Buca2, Marko1, BucaOTOC}.

The effect of bath on disorder-induced MBL has been studied, both experimentally and theoretically, showing stretched exponential decay of observables \cite{Open_disorder_MBL_PhysRevX_exp, Open_disorder_MBL_PhysRevLett_th1, Open_disorder_MBL_PhysRevLett_th2, Open_disorder_MBL_PhysRevE_th1}. SMBL also exhibits a stretched exponential decay of population imbalance with time, in the presence of interaction when the system experiences dephasing \cite{Open_SMBL_Eckardt_PhysRevLett}. However, an important aspect of the SMBL dynamical {\sl $\ell-$bits} remains unanswered, viz., what happens to them when the system is coupled to external (thermal) baths (cf. \cite{QSynch,Hadiseh})? 

Here we answer this question by focusing on the chain of interacting electrons in a tilted magnetic field gradient. Crucially, the experiments performed on cold atoms and trapped ion set-ups \cite{Scherg, morong2021observation} cannot shed light on the effects of a thermal environment, such as phonons, because phonons are essentially absent in optical lattices. Therefore, we consider a prototypical system of exchange-coupled quantum dot (QD) arrays where phonons appear quite naturally. Not only have the solid-state QD arrays shown promise for quantum-simulating various physical phenomena ranging from quantum information and computation \cite{Loss_DiVincenzo_PhysRevA, Zanardi_PhysRevB, Zanardi_PhysRevLett, Germanium}, Fermi-Hubbard model and Mott physics \cite{Byrnes_PhysRevB2008, Yang_PhysRevB2011, Hensgens2017}, Heisenberg magnetism \cite{vandyke2020protecting, Takumi_APL_2018, Mukhopadhyay_APL_2018, Mills2019, Qiao_PhysRevX_2020, Sigillito_PhysRevApplied2019}, Kondo physics, superconductivity \cite{Barthelemy_2013, Kouwenhoven_2001}, to even time-crystallinity \cite{Qiao2021, Economu_PhysRevB_2019, sarkar2021signatures, Sarkar_2022gex, Sacha, Seibold, Zhao2, Bakker}, but they are also sufficiently tunable in terms of bath engineering \cite{Granger2012, Mills2019}. For example, a reasonably tunable phonon bath can be achieved via metallic quantum-point contacts (QPCs), which in turn act as charge measuring apparatus \cite{Granger2012}.

Here we show, by focusing on the prototypical example of interacting spins in a magnetic field gradient, that dynamical {\sl $\ell-$bits} and the corresponding SMBL, survive when the system is coupled to local phonon baths at each site. For specific calculations, we consider an exchange-coupled quantum dot (QD) array connected to local phonon baths. The energy scale considered here has been achieved in recent experiments \cite{Byrnes_PhysRevB2008, Yang_PhysRevB2011, Hensgens2017, vandyke2020protecting,Takumi_APL_2018,Mukhopadhyay_APL_2018,Mills2019,Qiao_PhysRevX_2020,Sigillito_PhysRevApplied2019}.

The central premise of our results is that Stark/disorder-free many-body can be observed even in the presence of dissipations in the form of phonons, and crucially, this localization can protect the quantum coherence of a logical qubit for essentially infinite time. This immediately opens up a completely new conceptual and fundamental mechanism of implementing logical qubits in nano-scale quantum dot array set-ups and has far-reaching implications for quantum information processing and error corrections. One of the main significance of our results is that our new form of quantum coherence protection goes beyond standard decoherence-free and noiseless subsystems. This is of exceptional importance to the topics of \emph{quantum information in the context of many-body physics}, \emph{quantum gates}, and \emph{open quantum systems} because it offers a conceptually novel technique for protecting quantum coherence.

However, our proposal is not free from possible caveats. Successful implementation of SMBL in a QD array set-up needs to address these caveats. Although implementing a magnetic field gradient (e.g., in $z-$direction) has been achieved experimentally \cite{essler_frahm_2005, Hensgens2017, Kotlyar_PhysRevB, Zanardi_PhysRevLett, Zanardi_PhysRevB}, suitably fast switching of the tilt field to $x-$direction can be non-trivial. It can be a limitation for implementing the Pauli$-X$ gate that we propose in section \ref{Sec:gates}. Another possible caveat is the presence of a global phonon bath. Although the SMBL survives in the presence of local phonon baths, they do indeed disappear in the presence of a global phonon bath (cf. \cite{Jamir1, Jamir2}). In the former case, the correlation functions exhibit persistent oscillation with frequencies given by the dynamical {\sl $\ell-$bits}. In the latter case, the same correlation functions oscillated with a vanishing magnitude. However, at the very low operating temperature of semiconductor QD arrays, typically less than 50 mK, very few phonon modes of the host material, such as Si/GaAs, are excited. This leads to a weak electron-phonon and hence, spin-phonon interaction. Further reduction of electron-phonon scattering is possible, e.g., via interconnected oxide layers \cite{uno2005electron}. In short, overcoming the caveat corresponding to global phonon baths is not insurmountable but certainly non-trivial.

The paper is organized as follows. Sec.  \ref{Sec:model_setup} describes the model and how it can be implemented in a quantum-dot array set-up. Sec. \ref{Sec:auto_corr} shows transverse auto-correlation function as a function of time demonstrating the coherence of the dynamical $\ell-$bits. Secs. \ref{Sec:l-bit} and  \ref{Sec:l-bit_numerical_construct} show the mathematical and numerical constructions of the dynamical $\ell-$bits, respectively. Sec. \ref{Sec:l-bit_higher_spin} describes the higher spin generalization of the dynamical $\ell-$bits, and in Sec. \ref{Sec:conc} we conclude and discuss the importance of these novel dynamical $\ell-$bits.

\begin{figure*}
\includegraphics[keepaspectratio=true,scale=0.42]{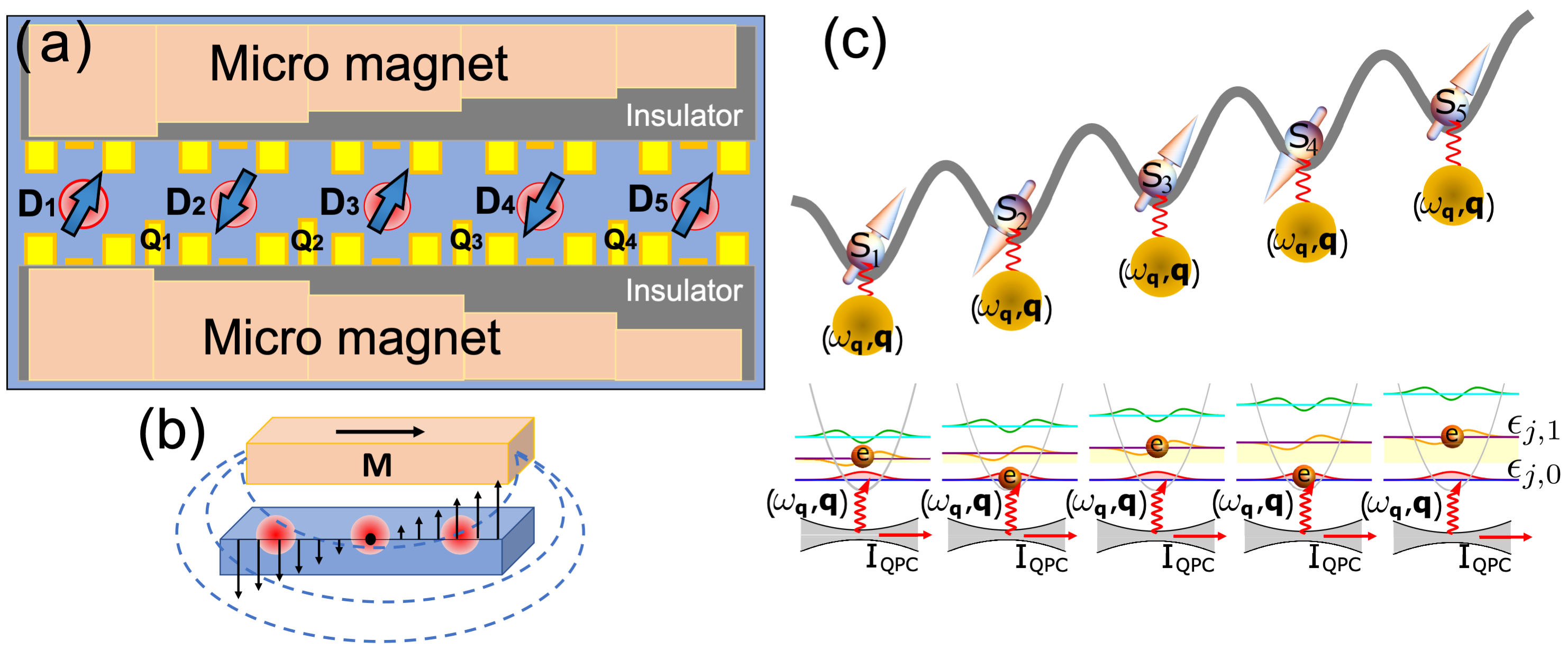}
    \caption{ {\bf Model and set-up:} (a) A  Schematic picture of a possible (experimental) setup of an array of 5 gate-controlled QDs, where $D_j$'s are the QDs and $Q_j$'s are the quantum point contact for proximate local phonons. (b) Magnetization of suitably placed micro-magnets produces a slanting magnetic field that affects the QDs (red blobs). (c) Schematic diagram of a tilted Heisenberg chain where each spin $\mathbf{S}_{j}$ is coupled to a local phonon-bath $(\hbar\omega_{\mathbf{q}}, \mathbf{q})_{j}$, with a possible realization of $H$ in semiconductor-based QD array set-up, where current through quantum point contact, $I_{\text{QPC}}$ would induce local phonons near the dot confinement \cite{Granger2012}.
    }
    \label{fig_set_up}
\end{figure*}

\section{Model and set-up}\label{Sec:model_setup}
We consider the paradigmatic SMBL spin$-S$ Hamiltonian on a chain of $N$ sites subject to coupling to a phonon bath, viz., $H = H_{s}+ H_{sb}$ where the system Hamiltonian is given by,
\begin{eqnarray}\label{spin_ham}
H_{s} &=& \sum_{j=1}^{N-1} J S_{j}\cdot S_{j+1} + \sum_{j=1}^{N} j W S_{j}^{z},
\end{eqnarray}
where $J>0$ represents that the spin-ground state is anti-ferromagnetic. Each site experiences a tilted magnetic field giving rise to Stark many-body localization. Moreover, each site is coupled to a local phonon bath
\begin{eqnarray}\label{sys-bath}
 H_{sb} &=& \sum_{j=1}^{N}\sum_{\mathbf{q}} \Big[ \hbar \omega_{j,\mathbf{q}}a_{j,\mathbf{q}}^{\dagger}a_{j,\mathbf{q}} + \big( \lambda_{j,\mathbf{q}}^{\perp} [S_{j}^{+} a_{j,\mathbf{q}} \nn \\ &+& S_{j}^{-} a_{j,\mathbf{q}}^{\dagger}] + \lambda_{j,\mathbf{q}}^{||} (a_{j,\mathbf{q}} + a_{j,\mathbf{q}}^{\dagger})S_{j}^{z}\big) \Big],
\end{eqnarray}
where $\hbar \omega_{j,\mathbf{q}}$ is the phonon energy,  $\lambda_{\mathbf{q}}^{\perp}$ is the spin-phonon coupling that can flip spins at each site and $\lambda_{\mathbf{q}}^{||}$ is the spin-phonon coupling that re-normalizes the energy-spectrum at each site by giving a shift in the $S^{z}-$ component of the spin. 

{Fig. \ref{fig_set_up} (a) shows the schematic set-up with the local phonon-bath and the tilted magnetic field produced by the micromagnets, motivated by Ref. \cite{Pioro-slanting2008}, \cite{jang2020robust}. A slanting Zeeman field can induce the Stark tilt field necessary for SMBL. Fig. \ref{fig_set_up} (b) shows how a micromagnet can produce a slanting magnetic field across a few quantum dot arrays. Such a magnetic field gradient can be implemented by embedding a micro-magnet of the appropriate geometry above or beneath the 2DEG \cite{Pioro-slanting2008, jang2020robust, Petta_Shuttling_PRB, Sigillito_PhysRevApplied2019}. Alternatively, specific to GaAs QDs, a suitable nuclear spin programming can create a large magnetic field gradient \cite{Zhang2017, Foletti2009, Bluhm_PhysRevLett}.}

{The semiconductor-based Quantum Dot (QD) simulator, therefore, supports the realization of Hamiltonian $H = H_{s} + H_{sb}$. In a semiconductor hetero-structure-based quantum dot the primary carriers are either electrons or holes \cite{HoleSpin1, HoleSpin2}. Therefore, a Fermi-Hubbard model is natural to arise in a QD simulator\cite{Hensgens2017, Kotlyar_PhysRevB, Zanardi_PhysRevLett, Zanardi_PhysRevB}. The confinement at each QD is approximately parabolic \cite{Kotlyar_PhysRevB} (i.e., a Harmonic oscillator), see Fig \ref{fig_set_up} (c). The magnetic field-dependent single particle energy at each dot corresponds to the standard Fock-Darwin-Zeeman scheme \cite{Kotlyar_PhysRevB}. Local electron-phonon interaction can be induced, e.g., by placing a biased Quantum Point Contact (QPC) \cite{Granger2012}, see Fig. \ref{fig_set_up} (c).} Having realized the Fermi-Hubbard model with a local carrier-phonon interaction, it can then be tuned to the desired limit to obtain a Heisenberg model (in the Mott limit) \cite{Hensgens2017, Kotlyar_PhysRevB, Zanardi_PhysRevLett, Zanardi_PhysRevB, essler_frahm_2005}. To this end, the carrier-carrier interaction strength, being related to the capacitance matrix of the QD array setup, can be tuned to reach a Mott limit \cite{Kotlyar_PhysRevB}.
Subsequently, the carrier-phonon interaction can also be mapped to the spin-phonon term in \eqref{sys-bath} (cf. \cite{Zanardi_PhysRevB, Zanardi_PhysRevLett}). 

The operating temperature of the quantum dot array (Si/GaAs) is usually a few tens of mili-Kelvin (mK) \cite{qiao2020conditional, mills2019shuttling}. At this temperature, the intrinsic phonons intrinsic phonons corresponding to the host materials, such as Silicon (deformation potential phonon), and GaAs (Polar optical phonon), are never local. However, these phonons have typical Debye temperature $\Theta_{D}$ of the order of a few hundreds of Kelvin, e.g., $\Theta_{D}  = 417~K$ in Si. Therefore, at 10 mK only a few phonon modes are excited. Further engineering can reduce the electron-phonon scattering strength \cite{uno2005electron}, effectively reducing the spin-phonon scattering. Moreover, since only a few phonon modes are excited at mK temperature regime the infinite-dimensional phonon Hilbert space can be truncated for numerical computation \cite{sarkar2020environment}. The above discussion remains true for phonons coming from a nearby metallic quantum point contact with $\Theta_{D}~(227 ~K~\text{in Silver})$ of the order of a few hundred Kelvin.

\section{Results} \label{Sec:numerical_results}
Having described how the paradigmatic SMBL spin-$S$ Hamiltonian can be implemented in a nano-scale QD array set-up, we first numerically show that our model exhibits coherence in the presence of spin-phonon interaction. For our numerical calculations we consider the Heisenberg exchange coupling to be $J=10~\mu eV$ (it can be a few tens of $\mu eV$ as well) \cite{Petta_Science, Qiao_PhysRevX_2020, Qiao2021, vandyke2020protecting}. In the following, we normalize the exchange coupling to $J=1$ for all our numerical calculations so that all other parameters are in the units of $10~\mu eV$. We consider a Holstein phonons \cite{holstein_phonon1, Dee_Holstein_2020} with phonon energy $\omega_{j,q} = \omega_0$ and $\lambda_{j,\mathbf{q}_{0}}^{\perp} = \lambda_{j,\mathbf{q}_{0}}^{||} = \lambda_{0}$. In principle, the momentum dependence of the spin-phonon interaction matrix element should have been chosen but that would prohibit us from doing exact numerical calculation. However, in each QD only a selected few energy-momentum conserved phonons interact with the spins.

\subsection{Transverse auto-correlation}\label{Sec:auto_corr}
\begin{figure}
    \includegraphics[keepaspectratio=true,scale=0.42]{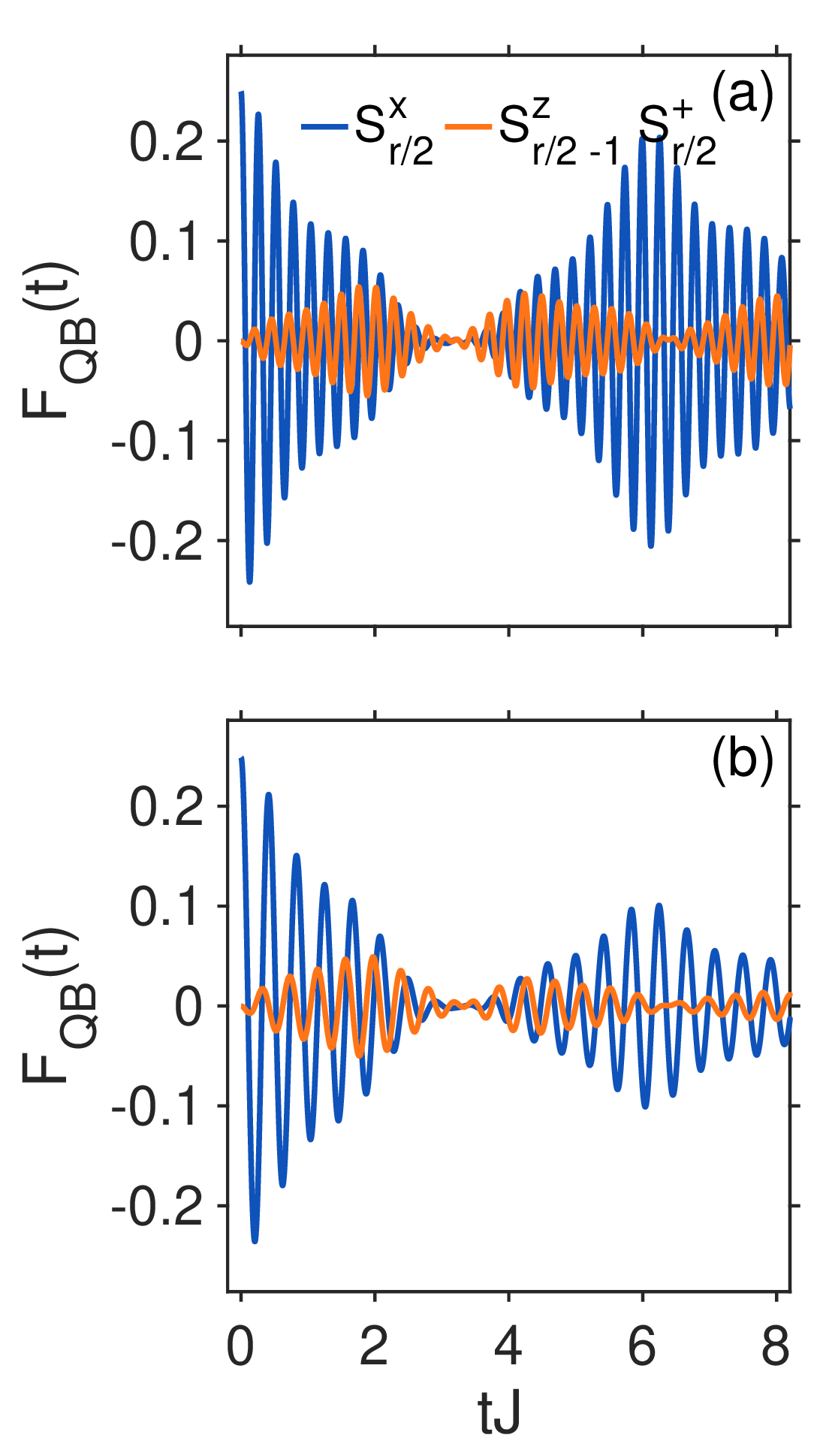}
    \caption{ {\bf Transverse auto-correlation:} $F_{QB}(t)$ for various choices of local operators $Q = S_{r/2}^{x}$ (blue line) and $S_{r/2 -1}^{z}S_{r/2}^{+}$ (orange line) corresponding to the system parameters (normalized to some integer values for convenience) $J = \Delta = 1$, $\lambda_0 = J$, $\omega_0 = 3J$, and $r = (N+1)$. (a) $N = 7, N_B = 3, W =6$ and (b) $N = 9, N_B = 2, W =3$. The sudden perturbation at $t=0$ is considered to be $S_{r/2}^{x}$.}
    \label{Fig:autocorrelation}
\end{figure}

To show the effect of a sudden perturbation, we focus on the maximally mixed case, for which the relevant function from linear response theory is the fluctuation function, $F_{QB} (t) = \frac{1}{2}\langle \lbrace Q(t), B \rbrace \rangle = \langle Q(t), B \rangle$ \cite{Buca_l-bit}. Considering a transverse perturbation field $B = S_{j}^{x}$ at $t = 0$, Fig. \ref{Fig:autocorrelation} plots the auto-correlation function corresponding to local operators $Q = S_{r/2}^{x}$ and $S_{r/2 -1}^{z} S_{r/2}^{x}$ for two different system sizes (\(N\)) and boson dimensions (\(N_B\), truncated phonon Hilbert space), specifically $(N = 7, N_B = 3)$ and $(N = 9, N_B = 2)$, respectively. Physically, $\langle S_{r/2}^{x}(t), S_{r/2}^{x} \rangle$ signifies the transverse auto-correlation at site $r/2$, and $\langle [S_{r/2 -1}^{z} S_{r/2}^{x}](t), S_{r/2}^{x} \rangle$ represents a three-spin auto-correlation function, with the former being directly measurable.

Two cases show persistent many-body Bl\"och oscillations, where in case of $S_{r/2 -1}^{z}S_{r/2}^{x}$, significantly low amplitude oscillations (the orange curve) are observed. In this case, the time-evolved operator acts on sites away from the perturbation site, signifying the memory of the perturbation is highly localized in space. Comparison of Fig. \ref{Fig:autocorrelation}(a) and (b) further shows that an increased tilt field $W~(>J)$ increases both frequency and amplitude of the coherent oscillation (see SI Fig. 1 for comparison for a fixed system size $N = 7$), thereby, signifying the necessity of a large tilt field for the coherence to be protected. We further find that the tilt field determines both the frequency and amplitude of the coherent oscillations, larger phonon energy supports the coherence, and the magnitude of the tilt field must be the largest energy scale in the system. See Appendix \ref{App:size_and_others} for further explanations.

Moreover, $F_{QB} (t)$ is physically related to the time-dependent local transverse susceptibility for the chosen operators and perturbation. In fact, due to the dynamical $\ell-$bits a lower bound for the persistent oscillations of the susceptibility can be found \cite{Marko1, Marko2}.

\subsection{Dynamical l-bits in the presence of local phonons}\label{Sec:l-bit}
The oscillatory transverse auto-correlation $\langle  S_{r/2}^{x}(t), S_{r/2}^{x} \rangle$ in Fig.\ref{Fig:autocorrelation} indicates the presence of a persistent coherence in the spin-phonon system. To analyze the physical origin of the coherence, we further analyze the role of the Stark field gradient. We now show that within the model $H$, how SMBL can be used to obtain a dynamical {\sl $\ell-$bit} would be protected against the environment. The field gradient induces these $\ell-$bits, which for the full Hamiltonian should be composed of operators that are combinations of the basis operators, viz., $\lbrace I, S_{j}^{+}, S_{j}^{-}, S_{j}^{z} \rbrace$.

\paragraph*{}
It has been shown that in the limit of large tilt field $W$, the time-evolution of any local operator $\mathcal{O}$ is governed by an effective Hamiltonian $H_{s,\text{eff}} = Y (D + M) Y^{\dagger}+O(J^3/W^2)$ such that $e^{i H_s t} \mathcal{O} e^{-i H_s t} \approx e^{i H_{s,\text{eff}} t} \mathcal{O} e^{-i H_{s,\text{eff}} t}$, where $D$ is the projection of $H_s$ onto the space of $\displaystyle M = \sum_{j} j W S_{j}^{z}$ \cite{pre-therm,ElsePreth,Buca_l-bit}. Here $Y$ is a quasi-local unitary operator close to identity and $H_{s,\text{eff}}$ governs the time evolution up to an exponentially long time $t^{*} \propto e^{W}$ \cite{Buca_l-bit,pre-therm}. The effective Hamiltonian $(D + M)$ has been shown to exhibit dynamical $\ell-$bits \cite{Buca_l-bit}. As long as the phonons interact locally enough (i.e. not globally) with the spins at each site we show below that we can still find an operator $Y$ such that an effective Hamiltonian can be obtained that governs the time evolution \cite{Francisco}.

In our spin-phonon system, the polaron transformation can decouple spin and phonon degrees of freedom in the large $W$, leading to an effective spin-phonon Hamiltonian, $\displaystyle D = H_{ZZ} + \sum_{j} \left( \sum_{\mathbf{q}}|\lambda_{j,\mathbf{q}}^{||}|^{2}\right) (S_{j}^{z})^{2}$, see Appendix \ref{App:math_formulation} for the details of the calculations. Therefore, the effective spin Hamiltonian now becomes,
\begin{eqnarray}\label{eq:h_eff}
    H_{\text{eff}}^{\prime} = \sum_{j} \big[ J S_{j}^{z}S_{j+1}^{z} - g_j (S_{j}^{z})^{2} + W j S_{j}^{z}\big],
\end{eqnarray}
where $g_{j} = \sum_{\mathbf{q}}\frac{|\lambda_{j,\mathbf{q}}^{||}|^{2}}{\hbar \omega_{j,\mathbf{q}}} $. Therefore, in the large tilt limit (above which the SMBL appears) the only remaining effect of spin-phonon interaction is the appearance of a single ion anisotropy in the effective Hamiltonian. The effective Hamiltonian $H_{\text{eff}} = Y H_{\text{eff}}^{\prime} Y^{\dagger}$ governs the time evolution for an exponentially long time with an error which is exponentially small in $W$ in the sense of \cite{pre-therm}, i.e. there exists a quasi-local basis transformation that maps the physical Hamiltonian into the effective one and which generates the time evolution up to exponentially long times (see also \cite{prethermlecture} for a pedagogical overview). Note that in the high limit large $W$ the phonon momentum dependence of the spin-phonon coupling $\lambda_{j,\mathbf{q}}^{||}$ and the phonon dispersion, viz., $\omega_{\mathbf{q}}$ (i.e., acoustic or optical) do not play any role. This {\em a-posteriori} justifies the use of the Holstein phonon in the numerical calculation. 

In a spin$-\frac{1}{2}$ system the single ion anisotropy term $(S_{j}^{z})^2 $ in \eqref{eq:h_eff} become an additive constant $\sum_{j} g_{j}$. The spin$-\frac{1}{2}$ effective Hamiltonian $H_{\text{eff}}^{\prime}$ hosts four linearly independent dynamical $\ell-$bits for each site $j$ satisfying, $\displaystyle [H_{\text{eff}}^{\prime},A_k(j)]=\omega_k A_k(j)$, viz., $ {A}_1(j) = {S}^+_j - 4{S}^z_{j-1}{S}^+_j{S}^z_{j+1},~
    {A}_2(j) = {S}^z_{j-1}{S}^+_j - {S}^+_j{S}^z_{j+1}, ~
    {A}_{3(4)}(j) = {S}^+_j \pm 2{S}^z_{j-1}{S}^+_j \pm 2{S}^+_j{S}^z_{j+1} + 4{S}^z_{j-1}{S}^+_j{S}^z_{j+1}$,
with four distinct frequencies per site,
 $ \omega_1(j) = \omega_2(j) = Wj,\ \omega_{3(4)}(j) = Wj \pm J$. Physically, these represent coherent localized excitations responsible for an oscillatory transverse auto-correlation function in Fig. \ref{Fig:autocorrelation}. 

  { Note that two of the dynamical $\ell-$ bits, viz., $A_{1(2)}(j)$ exhibits coherence frequency $Wj$ that is independent of the magnitude of the exchange interaction $J$. Therefore, any charge noise coming from various sources of the QD-array set-up that affects the exchange interaction $J \rightarrow J\pm \epsilon(t)$, where $\epsilon(t)$ is a time-dependent noise term, does not alter the frequencies of $A_{1(2)}(j)$.}

Physically, these represent coherent localized excitations. The quantum coherent nature of these excitations can be observed in the following way. The local symmetry ($\ell-$bit) of the model, $Q_k(j)=[A^\dagger_k(j),A_k(j)]$ satisfy $[H_{\text{eff}}^{\prime},Q_k(j)]=0$. Then for any equilibrium (not necessarily low temperature) state of the model $\rho_{eq}$, $[H_{\text{eff}}^{\prime},\rho_{eq}]=0$ implies $[Q_k(j),\rho_{eq}]=0$, and
 the states $A^\dagger_k(j)\rho_{eq}$ and $A_k(j)\rho_{eq}$ are finite frequency excitations that oscillate with frequency $\pm \omega_k(j)$. Crucially, these states are off-diagonal in the physical environmental (and many-body) eigenbasis of $Q_k(j)$ and therefore can store a qubit. This is reminiscent of  {noiseless subsystems} \cite{noiseless1,noiseless2,noiseless3} and other more exotic decoherence-free structures \cite{Albert,BucaAlgebraic}. However, in our case, there is no clear distinction between the coherent  {system} and decoherent {environment} because a dynamical $\ell-$bit centered around site $j$ spreads into other neighboring sites and has spatial overlap with neighboring dynamical $\ell-$bits (i.e. site $j$ cannot be thought of as being the system and the rest a bath). 

Since the effective Hamiltonian for the SMBL 
model is actually ${H}_{\rm{eff}}$ rather 
than ${H}_{\rm{eff}}'$, for the above 
dynamical $\ell-$bits ${A}_j(r)$, we can 
find that $\displaystyle [{H}_{\rm{eff}}, {Y}
{A}_j(r){Y}^\dagger] = {Y}[{H}_{\rm{eff}}', 
{A}_j(r)]{Y}^\dagger = \omega_j {Y}{A}_j(r)
{Y}^\dagger$. This implies ${Y}{A}_j(r)
{Y}^\dagger$ represents the set of dynamical 
$\ell-$bits corresponding to 
${H}_{\rm{eff}}$ that oscillate with 
frequencies $\omega_{j}$ .

\subsection{Numerical construction of dynamical l-bits}\label{Sec:l-bit_numerical_construct}

\begin{figure*}
    \centering
    \includegraphics[keepaspectratio=true,scale=0.25]{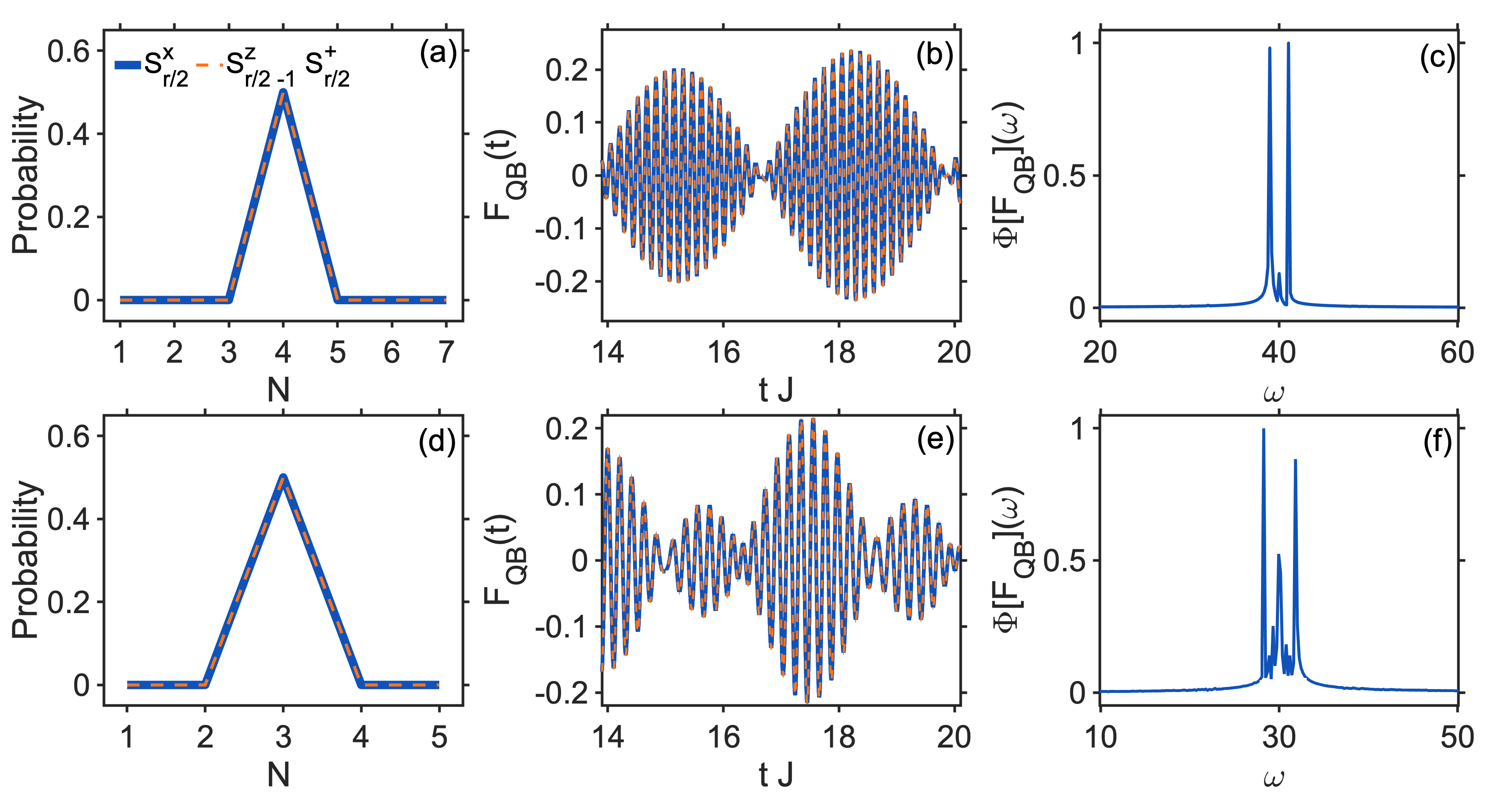}
    \caption{ {{\bf Numerical result for the dynamical $\ell-$bit candidate $\tau$:} The parameters used are $J = 1$, $W = 10$, and total time $T = 100$ to approximate the limit as $T \rightarrow \infty$. For a 7-dot system with $N_B = 2$:
    (a) Demonstrates the locality of the operator $\tau$; the probabilities measure how much of $\tau$ is localized on each site, using seed operators $A_1(r/2)$ (blue) and $A_4(r/2)$ (orange).
    (b) Shows the auto-correlation function $F_{QB}(t)$ with $Q = \tau$.
    (c) Presents the Fourier transform $\Phi[F_{QB}](\omega)$ for $A_1(r/2)$ (blue). For a 5-dot system with $N_B = 3$:
    (d) Demonstrates the locality of the operator $\tau$ similar to (a) with the same seed operators.
    (e) Shows the auto-correlation function $F_{QB}(t)$.
    (f) Presents the Fourier transform $\Phi[F_{QB}](\omega)$ for $A_1(r/2)$ (blue). The number of quantum dots chosen is consistent with spin-qubit arrays available in the QD array architecture.
} }
    \label{Fig:l-bit}
 \end{figure*}
 Despite the existence of dynamical $\ell-$bits for the full Hamiltonian $H$, we don't yet have them explicitly since we do not know the operator $Y$. In what follows we numerically show the dynamical $\ell-$bits. The auto-correlation in Fig.\ref{Fig:autocorrelation} indicates that the dynamical $\ell-$bits of the full Hamiltonian should be composed of operators combinations of the basis operators, viz., $\lbrace I, S_{j}^{+}, S_{j}^{-}, S_{j}^{z} \rbrace$. To this end, following Ref. \cite{Buca_l-bit, Mierzejewski_PhysRevLett} we numerically construct a dynamical $\ell-$bit operator, $\displaystyle \tau(j) = \lim_{T\rightarrow\infty} \int_{-T}^{T} dt~ e^{-i\omega_{k} t} U^{\dagger}(t) A_{k}(j) U(t)$, where the time evolution operator $U(t) = e^{- i H t}$ corresponds to the full (electron+phonon) Hamiltonian. Fig. \ref{Fig:l-bit} plots the numerical results corresponding to the properties of the dynamical $\ell-$bit operator $\tau(j)$ for a tilted field $W = 10J$. Following \cite{Buca_l-bit} we determine the locality of $\tau(j)$ by evaluating its components in a complete orthonormal product operator basis made up of products of (appropriately normalized) single site basis operators $\lbrace I, S_{j}^{+}, S_{j}^{-}, S_{j}^{z} \rbrace$ and plot in \ref{Fig:l-bit}(a) for a $7$-QD system with $N_B = 2$. The localization probabilities of $\tau(j)$ make it clear that it is sufficiently (quasi-)local quantity for the choices of seed operators $A_{k}(j)$, which agrees with our hypothesis. Thus we have numerically found quasi-local dynamical $\ell-$bits of the full Hamiltonian $H$. The auto-correlation function $F_{AB}(t)$ corresponding to $A = \tau(j)$ and $B=S_{j}^{x}$ indicates more than the one expected frequency comes into the dynamics in Fig. \ref{Fig:l-bit}(b) and also seen from the Fourier transform of the auto-correlation $\Phi[F_{AB}](\omega)$ plotted in Fig. \ref{Fig:l-bit}(c). We conjecture that this is due to the dynamical $\ell-$bit being spatially broader with the phonon than without the phonon. This implies that the dynamical $\ell-$bit excitation is reflected from the edge of our small chain giving two frequencies one above and one below the main one. This would be consistent with dynamical $\ell-$bits surviving the phonon bath, but being slightly less localized. The above conclusions hold for a $5$-QD system with $N_B = 3$ too, as seen from \ref{Fig:l-bit}(d)-(f). Moreover, comparing \ref{Fig:l-bit}(c) and (f) we find that with higher values of $N_B = 3$, the main frequency is becoming prominent. This indicates that with more QDs in the QD array set-up and with higher values of $N_B$ the expected main frequency should appear alone, corroborating our analytical argument. {However, increasing \(N_B\) without increasing chain-length \(N\) would develop noise around the central frequency, see Appendix \ref{App:size_and_others}.}
  
  \subsection{Thermalistion based quantum gate correction }\label{Sec:gates}
   We now show that the dynamical $\ell-$bits can be used to implement self-correcting gates where correction comes from thermalization itself. It will be convenient to consider the problem in the Heisenberg picture, i.e. study the time evolution of operators. We consider $A_{2}(j)$ as our main working $\ell-$bit operator and notice that $Q_2 (j)$, $A_{2}(j)$ and $A_{2}^{\dagger}(j)$ satisfy SU(2) algebra, viz., $Q_2 (j)=[A^\dagger_2(j),A_2(j)]$, $[Q_2 (j), A_2(j) ]= - 8 A_2(j)$ and $[Q_2 (j), A_2^{\dagger}(j) ] = 8 A_2^{\dagger}(j)$. Therefore, we can consider $Q_2 (j)$, $A_{2}(j)$ and $A_{2}^{\dagger}(j)$ as equivalent to the effective spin $\Sigma^{z}_{j}$, $\Sigma^{-}_{j}$ and $\Sigma^{+}_{j}$, respectively, in the space of dynamical $\ell-$bits. $H_{\text{eff}}^{\prime}$ acts as a rotation gate $R_{\Sigma^{z}}(t) = e^{iH_{\text{eff}}^{\prime} t}$ in the sense that it operates on our logical qubits as $R_{\Sigma^{z}}^{\dagger}(t) Q_2 (j) R_{\Sigma^{z}}(t) = Q_2 (j)$, and $R_{\Sigma^{z}}^{\dagger}(t) A_2 (j) R_{\Sigma^{z}}(t) = e^{i W j t} A_2 (j)$ which adds overall phase. The effect of this gate can be removed by considering the dynamical l-bits in their co-rotating bases at frequencies $W j$. Hence, we may consider this gate as the default one and store quantum information during its operation. We call this default application of $R_{\Sigma_{z}}$ the \emph{holding phase}, applications of other gates will be called simply the \emph{gate phase}.

To consider how the errors are corrected we first consider a \emph{generic} error $E$ for which $[M, E]\neq0$. Such an error will only be apparent after an exponentially long time scale as per the arguments in \cite{Buca_l-bit,pre-therm}. Only errors of the type $[M, E]=0$ are potentially relevant. Consider such a random error destroying the effective spin $\ell$-bits $\Sigma^\alpha_{j} \to \Sigma^\alpha_{j} + \varepsilon$. {Generically, we consider errors orthogonal to the dynamical $\ell$-bits, $\ave[\varepsilon \Sigma^\alpha_{j}]=0$. Such errors will not change $\ave[\Sigma^\alpha_{j}]$ and the thermalization dynamics will restore the dynamical $\ell$-bits. To see this, consider the dynamics again in the Schrodinger picture. Consider the initial state to be a time-dependent generalized Gibbs ensemble,  $\rho_{\text{tGGE}}$ corresponding to $H_{\text{eff}}^{\prime}$ \cite{BucaPRX}. This $\rho_{\text{tGGE}}$ is reached by the thermalisation dynamics of $H_{\text{eff}}^{\prime}$. Following the appearance of error, this state will change but the expectation values of the dynamical $\ell$-bits will not. This is because the long-time thermalization dynamics and corresponding thermalized state are fully determined only by the expectation values of dynamical symmetries (i.e., the dynamical $\ell$-bits) and conservation laws \cite{BucaPRX}. Therefore, the state as before the error has occurred will be restored by the thermalization dynamics of $H_{\text{eff}}^{\prime}$ \cite{ETHReview}. In the Heisenberg picture, this implies that the same operators $\Sigma^\alpha_{j}$ are restored.} Importantly, this argument holds irrespective of the error occurring during the gate phase or the holding phase.
 
  Before implementing an Ising gate with our logical qubits we note that an operator, defined on four adjacent lattice sites, $Q_2 (j) \otimes Q_2 (j+1)$ commutes with $H_{\text{eff}}^{\prime}$ and therefore, can be added to the latter without hampering the dynamical symmetry condition. {Therefore, the rotation operator $R_{\Sigma^{z} \Sigma^{z}}(t) = e^{\frac{i}{2} t Q_2 (j) \otimes Q_2 (j+1)}$ functions as an `Ising- gate'. To implement such a gate in a quantum dot (QD) array, tunable inter-dot exchange coupling is required, extending up to second-nearest neighbor sites. The effective Ising part of the Hamiltonian, $H_{\text{eff}}^{\text{Ising, Q}} = Q_2 (j) \otimes Q_2 (j+1) $ \footnote{Note that this new effective Ising Hamiltonian is constructed from the dynamical $\ell-$bit operators.} facilitates free evolution by commuting with $H_{\text{eff}}^{\prime}$ for a duration $t$. This evolution is particularly designed to mitigate systematic errors in controlled phase gates, as detailed in \cite{Jones_PhysRevA_Ising}.} The same arguments for passive error correction as the ones discussed for the holding phase apply during the application of this gate. 

   {We now discuss the implementation a rotation gate around $x-$ axis equivalent to Pauli$-X$ gate using our dynamical $\ell-$bit operators.} 
   {A perfect such $R_{\Sigma^{x}}(t)$ would do $R^\dagger_{\Sigma^{x}}(t) \Sigma^{z}_{j} R_{\Sigma^{x}}(t)=\Sigma^{y}_{j}=i(A_2^\dagger (j) - A_2 (j))$. Due to the specific nature of our setup, such a gate is not easy to implement robustly, but, fortunately, we do not need to implement it robustly as thermalization will correct any ``errors" in a gate as we will now discuss.}  {In an ideal situation we add $\Sigma^{x}_{j} = A_2 (j) + A_2 ^{\dagger}(j)$ to the original Hamiltonian \eqref{spin_ham}. However, due to the large tilt field in the z-direction such a term would not appear in the effective Hamiltonian $H_{\text{eff}}$. If we now switch off the tilt field itself we then  $\Sigma^{x}_{j}$ term would remain in $H_{\text{eff}}$ and in the system albeit with reduced amplitude, but the $\ell-$ bits would start decaying on a time scale $J^{-1}$. Within this time scale an $\displaystyle H_{\text{eff}}(W=0) = \sum_{j} \big[ J S_{j}^{z}S_{j+1}^{z} + \Sigma^{x}_{j} \big]$ can implement an $X-$gate, viz.,  $R_{\Sigma^{x}}(t) = e^{i t H_{\text{eff}}(W=0) }$. Once the operation of the $X-$gate is done we can re-introduce the tilt field in the $ z-$ direction, and any intermittent error will be corrected by the thermalization.} 
   {The Ising gate and the $R_{\Sigma^{x}}(t)$  together can implement a CNOT gate \cite{PreskillReview}.}

\subsection{Generalization to higher spin systems}\label{Sec:l-bit_higher_spin}
With the dynamical $\ell-$bits for the spin$-\frac{1}{2}$ system at hand, we search dynamical $\ell-$bits for a general spin$-S$ system. The motivation behind this is not only theoretical but also the availability of spin$-\frac{3}{2}$ (heavy-) hole-based quantum dots \cite{Li_acs.nanolett_heavy_hole_qubit}. In this case, a numerical construction is more unlikely to be achievable. We therefore, concentrate on obtaining an analytical form for $A_{k}(j)$ and the corresponding effective Hamiltonian that would give us the seed operator, instead of finding $\tau(j)$. The single-ion anisotropy term in \eqref{eq:h_eff} is no longer an additive constant (i.e., a constant time Identity matrix). We consider (experimentally available) spin-$\frac{3}{2}$ case. Each site exhibits pairs of states ($|\frac{3}{2}\rangle$,$|-\frac{1}{2}\rangle$) and ($|-\frac{3}{2}\rangle$,$|\frac{1}{2}\rangle$) representing sectors that are decoupled because single-ion anisotropy can't couple them \cite{Oitmaa_PhysRevB}. The each of these sectors can be mapped to a set of Pauli operators, which for the sector ($|\frac{3}{2}\rangle$,$|-\frac{1}{2}\rangle$) is given by, $\displaystyle S_{j}^{z} \rightarrow \sigma_{j}^{z} +\frac{1}{2}$, to get a spin$-\frac{1}{2}$ Hamiltonian \cite{Oitmaa_PhysRevB}. Therefore, the effective Hamiltonian \eqref{eq:h_eff} turns out to be,
 \begin{eqnarray}\label{spin-s-h_eff}
 \tilde{H}_{\text{eff}}^{\prime} &=& \sum_{j} \bigg[ J \sigma_{j}^{z} \sigma_{j+1}^{z} - g_j (\sigma_{j}^{z})^{2} + (Wj +J -g_j) \sigma_{j}^{z}\bigg] \nn \\ &+& \text{const.},
 \end{eqnarray}
 where $\displaystyle \text{const.} = \left(\frac{J}{4} + j\frac{W}{2} - \frac{g_{j}}{4} \right)$. Comparison of \eqref{eq:h_eff} and \eqref{spin-s-h_eff} clearly shows that the structure of the dynamical $\ell-$bit operators $A_{k}(j)$ remains the same as that corresponding to the spin$-\frac{1}{2}$ system but, the corresponding frequencies get renormalized, viz., $ \tilde{\omega}_1(j) =\tilde{\omega}_2(j) = Wj - g_j, \tilde{\omega}_3(j) = Wj -g_j + J,  \tilde{\omega}_4(j) = Wj -g_j - J.$
Similar generalization, following \cite{Oitmaa_PhysRevB}, for other spin values can be obtained case by case.

\section{Conclusions and discussion}\label{Sec:conc}
Accessible and scalable quantum coherence can function as the basis of a quantum computer (qubits). In semiconductor set-ups, a major limiting factor for scalability is charge noise coming from the many-body interactions and phonon interactions which is the case we studied \cite{Germanium}. In this manuscript, we propose a novel way to protect quantum coherence in semiconductor structures, considering a one-dimensional model of a strongly interacting system that has been realized in semiconductor QD array set-up \cite{Hensgens2017, Mills2019, Sigillito_PhysRevApplied2019, Qiao_PhysRevX_2020, Qiao2021}.

We show analytically and numerically that a strong external magnetic field induces Stark many-body localization even in the presence of a local phonon bath that is a realistic source of dissipation. We identify a super-extensive number of spatially localized finite-frequency excitations in the system (dynamical $\ell-$bits). {The super-extensivity appears in the following way. Any product of neighboring 4 (per site) dynamical $\ell$-bits $A_{k_1}(j)A_{k_2}(j+x_1)A_{k_3}(j+x_2) \ldots$ is another dynamical $\ell$-bit that is linearly independent from the original dynamical $\ell$-bits $A_{k}(j)$. There is a super-extensive number of such dynamical $\ell$-bits.} 

The time scale for which the dynamical $\ell-$bits survive exhibits an exponential dependence on the strength of the tilt field ($t^{*} \propto e^{W}$). This implies a persistent oscillation in the auto-correlations functions at any temperature, and we numerically show this for the maximally mixed state. For small system sizes that are numerically accessible, we find an additional frequency in the dynamics that our analytical theory can't explain. We conjecture that this is due to the phonon bath inducing the dynamical $\ell-$bits to slightly spread in space. 

Crucially, in the $\ell-$bit (conservation law) basis of the system, the dynamical $\ell-$bit excitations are off-diagonal and hence have the potential to store quantum information as logical qubits. The dynamical $\ell-$bits are fully protected both from the many-body electron-electron interactions and the phonons coming from the quantum point contacts. In this sense, they provide a possible route to preserve quantum coherence in experimentally achievable mesoscopic semiconductor QD array set-up \cite{Heinrich2021, vandyke2020protecting, Petta_Science, Santos2020}. As the dynamical $\ell-$bits have overlap in space and are present throughout the system they, in a sense, extend standard notions of noiseless subsystems \cite{noiseless1,noiseless2,noiseless3} because there is no well-defined separation between the system and bath.

Our work thus opens the possibility of quantum error correction not only for quantum memory but also for gate operations \cite{PreskillReview}. More specifically, we have shown how to operate passively error-correcting gates on the dynamical $\ell$-bits that could allow for universal quantum computing in the setup proposed. The main advantage of our approach is that the thermalization dynamics of the Hamiltonian itself correct the errors that occurred both during the gate operation and during the storage. The setup should be contrasted with well-known noiseless subsystems, decoherence-free sub-spaces, and related structures \cite{Viola1,Viola2} because we do not require a clear physical separation between the system, hosting a quantum register, and an \emph{environment} that introduces errors. Rather in our case, the full many-body system is a register of mutually overlapping dynamical $\ell$-bits reinforced by the thermalization.

{Having emphasized the thermalization-based error correction, we briefly elaborate on the possible readout mechanisms. The go-to method for reading out semiconductor quantum-dot qubits is spin-to-charge conversion \cite{burkard_semiconductor_2023}. However, our proposal requires a dense array of many qubits, $\mathcal{O} \gg 10$. Then the spin-to-charge conversion readout requires two charge reservoirs to be placed near the qubit to be read, complicating the placement of the reservoirs in the dense array \cite{ciriano-tejel_spin_2021, oakes_fast_2023}. One way to circumvent this issue is a spin-readout method based on spin-dependent tunneling combined with gate-based reflectometry of a neighboring quantum dot to act as a charge sensor, as proposed in Ref. \cite{ciriano-tejel_spin_2021}. Similar single electron box-based readouts look promising for the readout \cite{oakes_fast_2023}.
}

\section{Acknowledgments}
SS acknowledges Yonatan Dubi for useful discussions. BB acknowledges funding from the EPSRC program grants EP/P009565/1, EP/P01058X/1, the EPSRC National Quantum Technology Hub in Networked Quantum Information Technology (EP/M013243/1) and was supported by a research grant (42085) from VILLUM FONDEN. 
\bibliography{main_1}

\appendix
\section{Dependence on system sizes, dimension of phonon Hilbert space, and system parameters}\label{App:size_and_others}

\begin{figure*}
    \centering
    \includegraphics[keepaspectratio=true,scale=0.36]{ 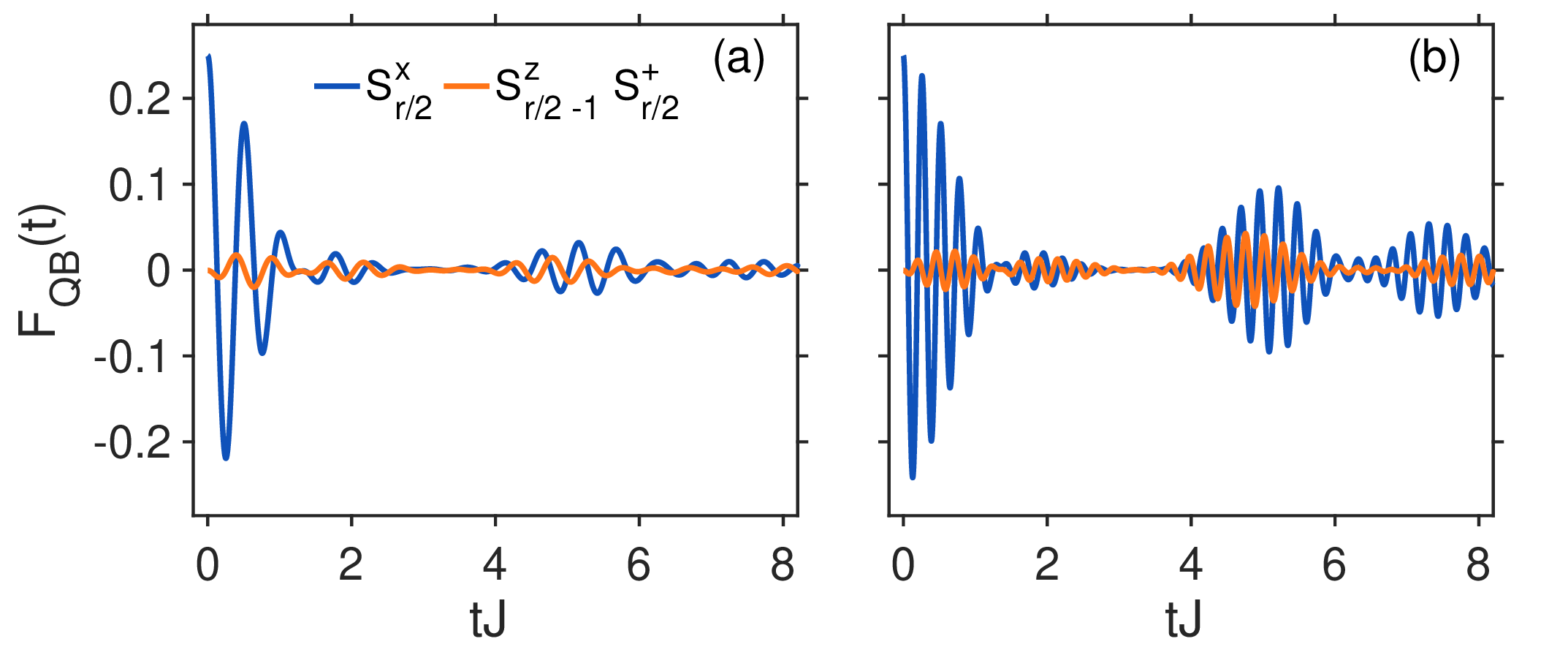}
    \caption{ {\bf Transverse auto-correlation for two different tilt field magnitudes for a fixed system size:} $F_{QB}(t)$ for local operators $Q = S_{r/2}^{x}$ (blue line) and $S_{r/2 -1}^{z}S_{r/2}^{+}$ (orange line) corresponding to the system parameters $N = 7, N_B = 3, J = \Delta = 1, \omega_0 = \lambda_0 = J$: (a) $ W = 3J$, (b) $W = 6J$, where $r = (N+1)$.}
    \label{Fig:autocor_vs_tilt}
\end{figure*}

 Fig. \ref{Fig:autocor_vs_tilt} plots the transverse auto-correlation function $F_{QB}(t)$ for two different values of tilt field for a fixed system size $N=7$ and fixed phonon Hilbert space truncation $ N_B = 3$. Comparing Fig. \ref{Fig:autocor_vs_tilt} (a) corresponding to $W = 3J$ and (b) corresponding to $W = 6J$ we find that for a fixed system size and dimension of the phonon Hilbert space ($N_B$) an increased tilt field $W~(>J)$ (in Fig. \ref{Fig:autocor_vs_tilt} (b))  increases both frequency and amplitude of the correlation.

\begin{figure*}
    \centering
    \includegraphics[keepaspectratio=true,scale=0.35]{ 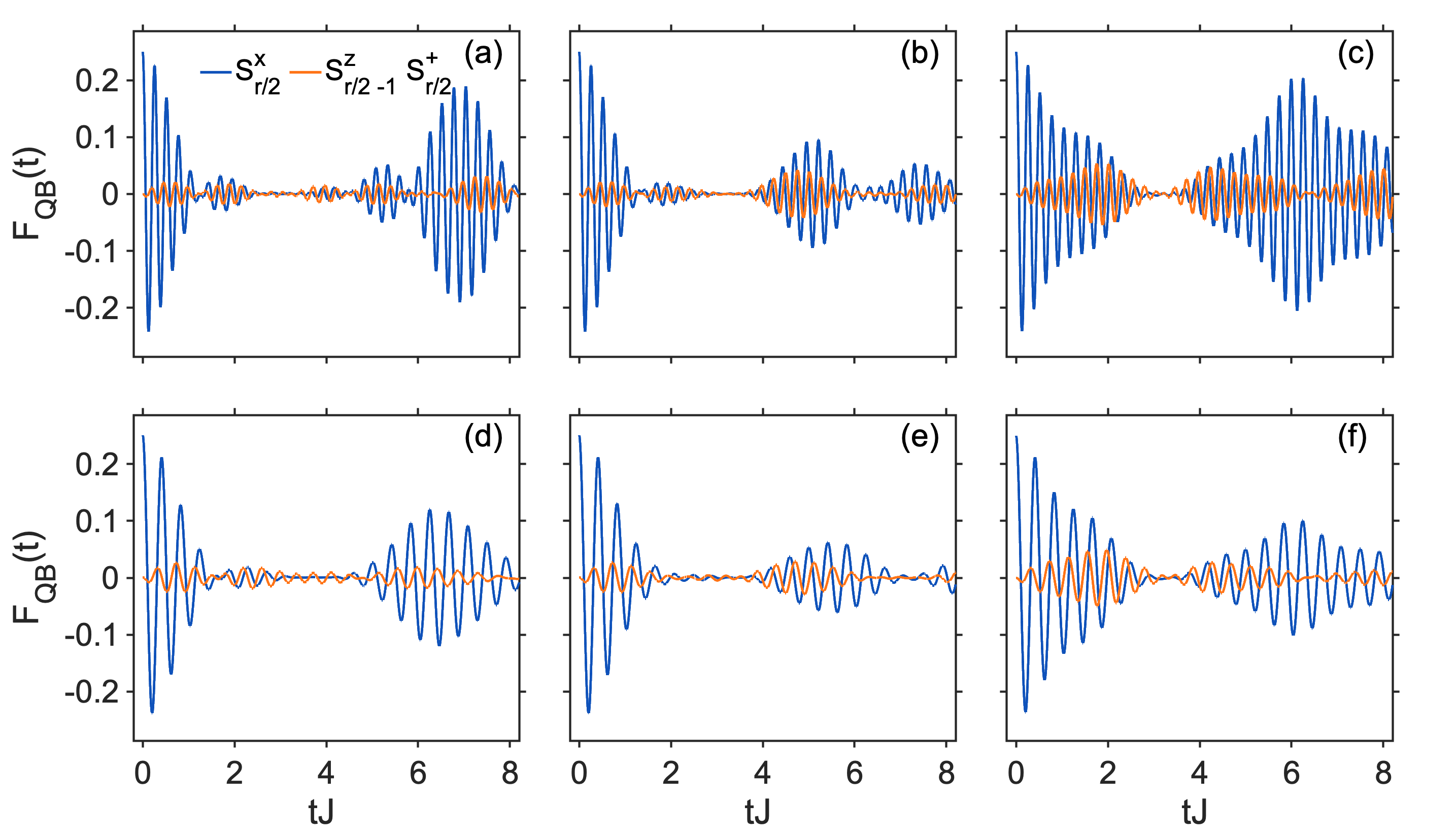}
    \caption{ {\bf Transverse auto-correlation for different system sizes and different values of phonon energies:} $F_{QB}(t)$ for various choices of local operators $Q = S_{r/2}^{x}$ (blue line) and $S_{r/2 -1}^{z}S_{r/2}^{+}$ (orange line) corresponding to the system parameters $J = \Delta = 1, \lambda_0 = J$: $N = 7, N_B = 3, W =6$, (a) $\omega_0 = J/3$, (b) $\omega_0 = J$, (c) $\omega_0 = 3J$; and $N = 9, N_B = 2, W =3$, (d) $\omega_0 = J/3$, (e) $\omega_0 = J$, (f) $\omega_0 = 3J$, where $r = (N+1)$. The sudden perturbation at $t=0$ is considered to be $S_{r/2}^{x}$, the same as in the main text.}
    \label{Fig:autocorr_vs_omega}
\end{figure*}
Fig. \ref{Fig:autocorr_vs_omega} plots the transverse auto-correlation function $F_{QB}(t)$ for different system sizes, different values of phonon energies, and dimensions of the phonon Hilbert space. Fig. \ref{Fig:autocorr_vs_omega}(a)-(c), corresponding to $N=7$ and $N_B = 3$, shows that for $\omega_0 > J$ oscillation in the auto-correlation function $F_{QB}(t)$ is more robust compared to  $\omega_0 < J$, despite persistent oscillation being seen across a large range of phonon energies. Fig. \ref{Fig:autocorr_vs_omega}(d)-(f), corresponding to $N=9$ and $N_B = 2$ also corroborate the same conclusions. Therefore, for larger phonon energies the oscillation in the transverse auto-correlation is more robust in terms of amplitude, which is expected because in the effective spin-poraronic description the effective polaron shift of energy spectrum is inversely proportional to the energy of the phonons, viz., $\displaystyle g_{j} = \sum_{\mathbf{q}} \frac{|\lambda_{\mathbf{q}}^{||}|^2}{\hbar \omega_{\mathbf{q}}} \approx \frac{|\lambda^{||}|^2}{\hbar \omega}$, note the resulting $\mathbf{q}-$independence comes from the fact that $|\lambda_{q}^{||}|^2 \propto |\mathbf{q}|$ (within the standard deformation potential approximation) and $\hbar \omega_{\mathbf{q}}\propto |\mathbf{q}|$ (due to linear phonon dispersion) too. For a fixed $\lambda_{\mathbf{q}}^{||}$ larger phonon energies reduces the polaron shift $g_j$.

\begin{figure*}
    \centering
    \includegraphics[keepaspectratio=true,scale=0.25]{ 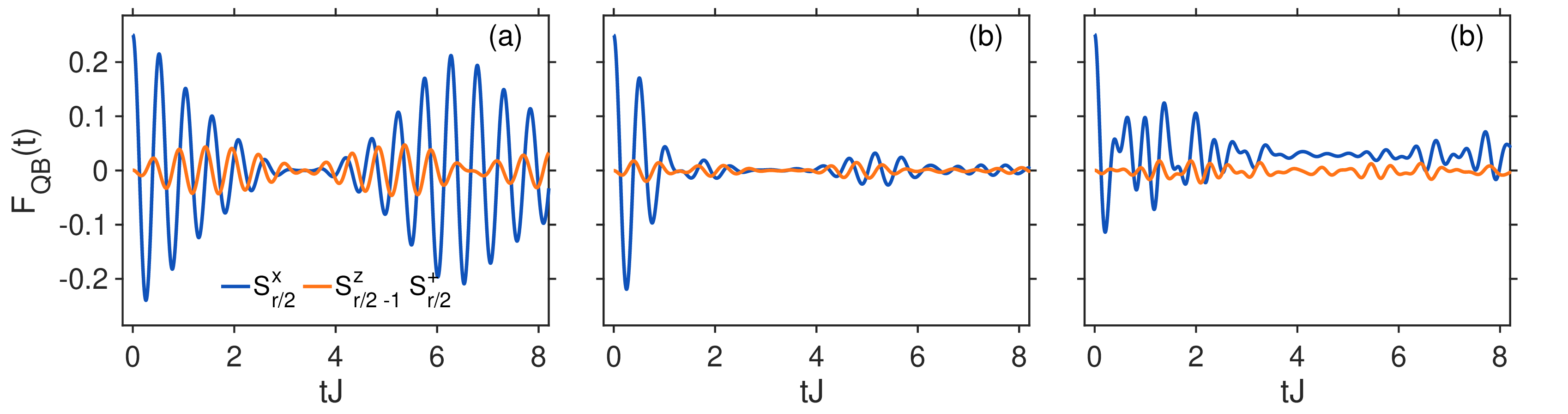}
    \caption{ {\bf Transverse auto-correlation for two different values of $\lambda_0$:} $F_{QB}(t)$ for local operators $Q = S_{r/2}^{x}$ (blue line) and $S_{r/2 -1}^{z}S_{r/2}^{+}$ (orange line) corresponding to the system parameters $N = 7, J = \Delta = 1, \omega_0 = J, W = 3J$: (a) $ \lambda_0 = J/3$, (b) $ \lambda_0 = J$, and (c) $ \lambda_0 = 3J$, where $r = (N+1)$.}
    \label{Fig:autocorr_vs_lambda}
\end{figure*}

Fig. \ref{Fig:autocorr_vs_lambda} plots the transverse auto-correlation function $F_{QB}(t)$ for three different values of spin-phonon coupling strengths $\lambda_0 = J/3$, $J$ and $3J$. Comparing Fig. \ref{Fig:autocorr_vs_lambda}(a)-(c) we find that for $ \lambda_0 \geq W$ the auto-correlation function $F_{QB}(t)$ no longer exhibits coherent oscillation. Moreover, comparing Fig. \ref{Fig:autocorr_vs_lambda}(b) and (c) we find that the coherent oscillation of the auto-correlation is also destroyed in terms of their frequency and becomes completely random.  This indicates that the tilt field must be the dominant energy scale in the system for the localization (and therefore $\ell-$bits) to survive. 
In summary, we find that (i) the tilt field determines both the frequency and amplitude of the coherent oscillations, (ii) larger phonon energy supports the coherence, and (iii) the magnitude of the tilt field must be the largest energy scale in the system.

However, increasing $N_B$ and keeping $N$ fixed would induce more noise in the system effectively destroying the coherence. 

\begin{figure*}[t]
\centering
\includegraphics[keepaspectratio=true,scale=0.34]{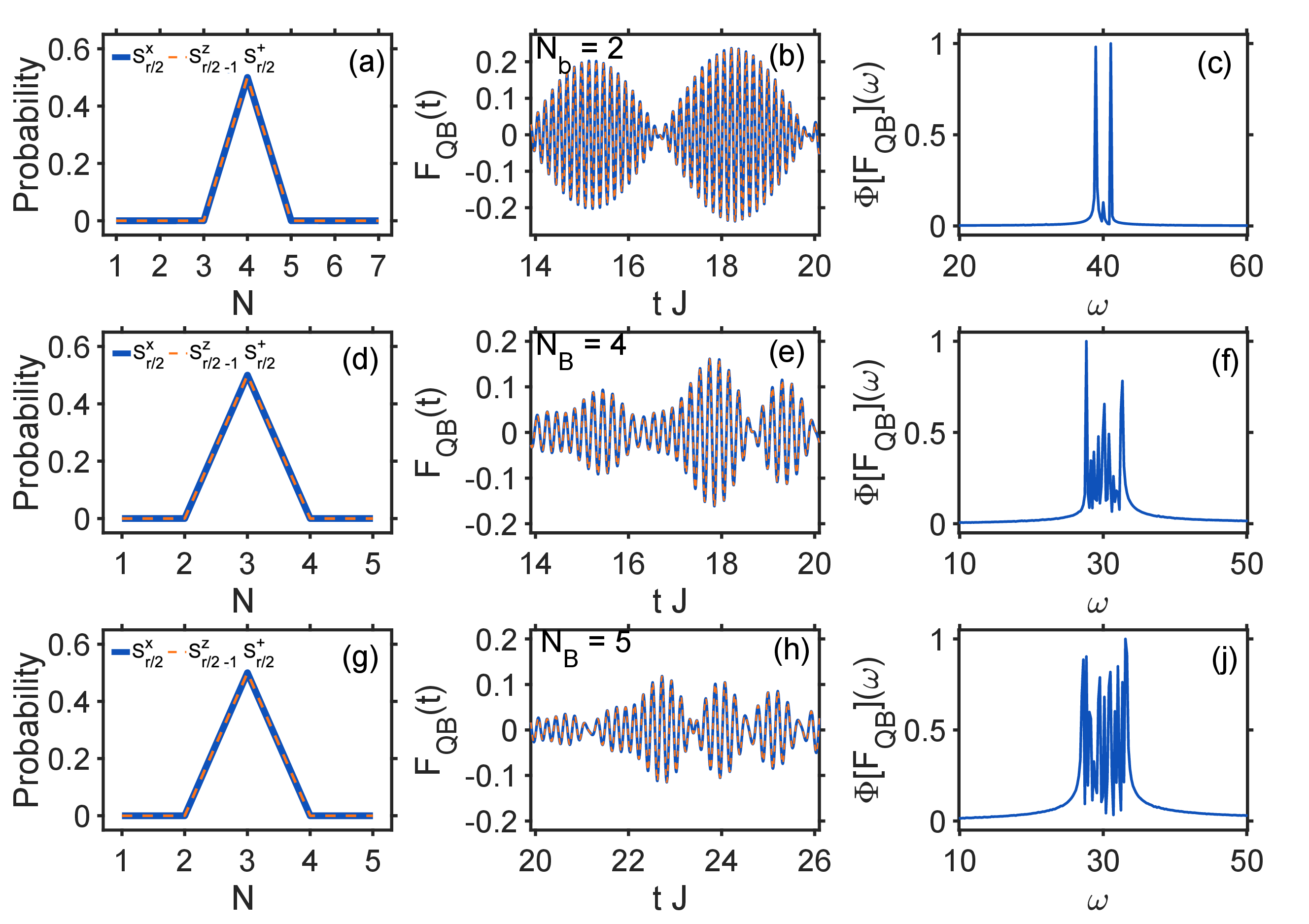}
    \caption{ {\bf Numerical result for the dynamical $\ell-$bit candidate $\tau$:} For a 7-dot system with $N_B = 2$, (a) - (c); For a 5-dot system with $N_B = 4$, (d) - (f); For a 5-dot system with $N_B = 5$, (d) - (f).  
}
    \label{Fig:l-bits_new}
 \end{figure*}
\begin{figure}[t]
    \centering
    \includegraphics[keepaspectratio=true,scale=0.35]{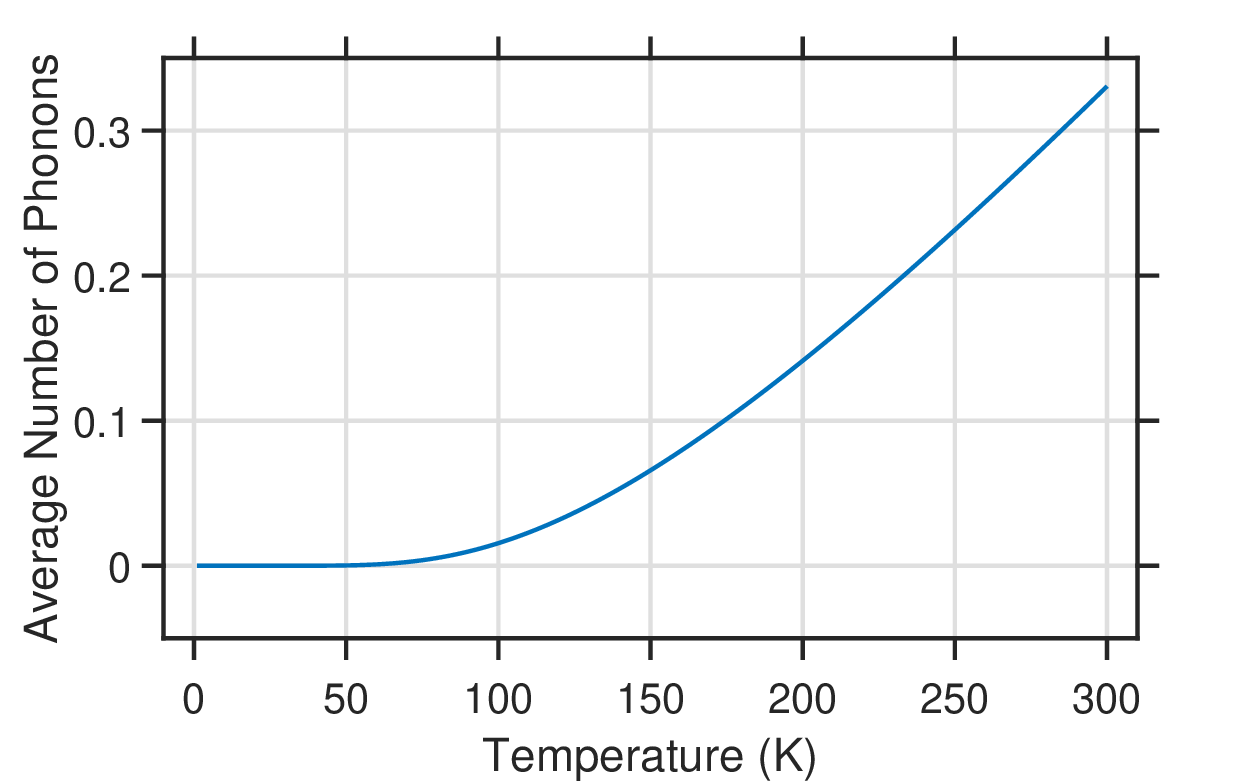}
    \caption{ {\bf Average number of phonons:} The average number of phonons, given by the Bose distributions, as a function of temperature. 
}
    \label{Fig:avg_phn}
 \end{figure}

From Fig. \ref{Fig:l-bits_new} it is clear that the value of $N_B$ would indeed affect the oscillation frequency by bringing in more noise. Still, the system size $N$ that is numerically achievable is limited to $N = 5$ when we consider larger values of $N_B$ because of the requirements of local phonons. However, as explained for a fixed $N_B$ with increasing $N$, the result would converge to the case corresponding to (a)-(c). Note that the locality of the variables $\tau$ given by the probabilities remains the same irrespective of the value of $N_B$. It is worthwhile to point out that in actual implementation in quantum dot array set-up, the physical operating temperature is usually in milli-Kelvin (cf. \ref{Sec:model_setup}), the phonons at thermal equilibrium obey the Bose-Einstein distribution:
\( \displaystyle
n(\omega_0) = \left(e^{\frac{\epsilon_0}{k_B T}} - 1\right)^{-1},
\)
where $\omega_0$ is the constant energy of the optical phonons, $k_B$ is the Boltzmann constant, $T$ is the temperature in Kelvin. For the optical phonons, we have considered an energy of \(\omega_0 \approx 36\) meV. Therefore, for most of the operating temperatures, the average number of phonons would be much less than what we have considered, i.e., $N_B = 2, 3,$ etc., see Fig. \ref{Fig:avg_phn}

\section{Mathematical formulation}\label{App:math_formulation}
 We note that $H_s = H_{XX} + H_{ZZ} + M$, where $H_{XX} = K + K^{\dagger}$ with $K = \sum_{j} \frac{J}{2} S_{j}^{+} S_{j+1}^{-}$, $H_{ZZ} = \sum_{j} J S_{j}^{z} S_{j+1}^{z}$ and $ M = \sum_{j} j W S_{j}^{z}$. Clearly the eigen-spectrum of $M$ is equally separated, i.e., successive eigenvalues of $M$ differ by $W$ the tilt field magnitude.

 As long as the phonons interact locally  (i.e. not globally) with the spins at each site we show below that we can still find an operator $Y$ such that an effective Hamiltonian can be obtained that governs the time evolution \cite{Francisco}. To show this we first denote different terms of Hamiltonian $H$ (Eq. (1) and (2) of the main text) in the following way: $H_{XX}^{sb} = \sum_{j} (h_{j}^{\perp} + h_{j}^{\perp \dagger}) $ with  $h_{j}^{\perp} = \sum_{\mathbf{q}} \lambda_{j,\mathbf{q}}^{\perp} S_{j}^{+} a_{j,\mathbf{q}} $, and $H_{ZZ}^{sb} = \sum_{j} h_{j}^{||} $ with  $h_{j}^{||} = \sum_{\mathbf{q}} \lambda_{j,\mathbf{q}}^{||} (a_{j,\mathbf{q}} + a_{j,\mathbf{q}}^{\dagger})S_{j}^{z}$. Using the fundamental commutation relation $[S_{i}^{z}, S_{j}^{\pm} ] = \pm S_{j}^{\pm} \delta_{i,j}$ one can easily show that,
\begin{eqnarray}\label{comm}
 \left[M, K\right] &=& -W K,~ \left[M, K^{\dagger}\right] = W K^{\dagger}, ~ \left[M, H_{zz}\right] = 0; \nn \\ 
 \left[M, h_{j}^{\perp}\right] &=& j W h_{j}^{\perp},~ 
 \left[M, h_{j}^{\perp \dagger}\right] = -j W h_{j}^{\perp \dagger},~ 
 \left[M, h_{j}^{||}\right] = 0. \nn \\
\end{eqnarray}
Then using the following form of $D$,

\begin{align}
    D &= \frac{1}{T}\int_{0}^{2\pi /W} dt e^{iMt} (H_s + H_{sb}) e^{-iMt} + H_{ph}, \nonumber \\
    &= \frac{1}{T}\int_{0}^{2\pi /W} dt \left[ e^{iMt}(K + K^{\dagger} + H_{zz}) e^{-iMt} \right] \nn \\ &+ \frac{1}{T}\int_{0}^{2\pi /W} dt \left[ e^{iMt} \sum_{j} (h_{j}^{\perp} + h_{j}^{\perp \dagger} + h_{j}^{||} ) e^{-iMt} \right] \nn \\ &+ H_{ph} \nonumber \\ &= \frac{1}{T}\int_{0}^{2\pi /W} dt \left[ e^{-iWt} K + e^{i Wt}K^{\dagger} + H_{zz}\right] \nn \\ &+ \frac{1}{T}\int_{0}^{2\pi /W} dt \left[ \sum_{j} (e^{i(jW)t}h_{j}^{\perp} + e^{-i (jW) t}h_{j}^{\perp \dagger} + h_{j}^{||} ) \right] \nn \\ &+ H_{ph} \nonumber \\ &= H_{zz} + \sum_{j} h_{j}^{||} + H_{ph}, 
\end{align}
we can easily obtain $D = H_{ZZ} + H_{ZZ}^{sb} + H_{ph}$, where $ H_{ph} = \sum_{j, \mathbf{q}} \hbar \omega_{j,\mathbf{q}}a_{j,\mathbf{q}}^{\dagger}a_{j,\mathbf{q}}$. We further show that the systematic elimination of terms that don’t conserve the dominant term $M$ order by order leads to a $1/W$ correction. We notice that $\displaystyle H = (M + K + K^{\dagger} + H_{zz} + \sum_{j}[h_{j}^{\perp} + (h_{j}^{\perp})^{\dagger} + h_{j}^{||}])$, following Ref. \cite{Buca_l-bit}. We write down the effective Hamiltonian as, \( H^{\prime} = e^{S} H e^{-S},\) where $S$ is a Hermitian operator and expand $S = \sum_{r} S^{(r)}$ with $S^{(r)} \sim \mathcal{O} (J^{r+1} / W^{r})$. Using the Backer-Campbell-Hausdorff relation, we group the terms in the expansion of $H^{\prime}$ order by order, 
\begin{align}
    H^{\prime} &= M + \underbrace{(H - M + [iS^{(1)}, M])}_{\mathcal{O}(1)} \nn \\ &+ \underbrace{([iS^{(1)}, H-M] + [iS^{(2)}, M])}_{\mathcal{O}(J^2/W)} + \mathcal{O}(J^3/W^2), \nonumber \\ &= M + \left(H_{zz} + \sum_{j} h_{j}^{||}\right) \nn \\ &+ \underbrace{([iS^{(1)}, H-M] + [iS^{(2)}, M])}_{\mathcal{O}(J^2/W)} + \mathcal{O}(J^3/W^2),
\end{align}
for choosing, $iS^{(1)} = \frac{1}{W} \left(K^{\dagger} - K + \sum_{j} \frac{1}{j} [h_{j}^{\perp} - (h_{j}^{\perp})^{\dagger}] \right)$. Note that, such a choice of $S^{(1)}$ keeps $S^{(1)}$ Hermitian. In the next order, we have to choose $iS^{(2)}$ such that $([iS^{(1)}, H-M] + [iS^{(2)}, M])$ at the best gives a $1/W$ correction. In the term \([iS^{(1)}, H-M]\), all the terms involving \([iS^{(1)}, K^{\dagger} + K + \sum_{j} [h_{j}^{\perp} + (h_{j}^{\perp})^{\dagger}]\) correspond to $\mathcal{O}(J^2 / W)$ and the rest of the terms must cancel with \([iS^{(2)}, M]\). Therefore, we choose \(iS^{(2)} = \frac{1}{W} [iS^{(1)}, (H_{zz} + \sum_{j} h_{j}^{||})]\). Therefore, the effective Hamiltonian becomes, 
\begin{align}
    H^{\prime} &= M + \left(H_{zz} + \sum_{j} h_{j}^{||}\right) + \frac{J^2}{ 2 W} (S_{N}^{z} - S_{1}^{z})\nn \\ &+ \sum_{j} 4 \frac{\sum_{ \mathbf{q}}|\lambda_{j,\mathbf{q}}|^2 }{jW} S_{j}^{z} + \text{other~}\mathcal{O}(J^2 / W).
\end{align}
\paragraph*{}
The leading order correction is $\mathcal{O}(J^2 / W)$ which is negligible in the large tilt field limit ($W>>J$). Indeed the referee is correct that the term $\sum_{j} 4 \frac{\sum_{ \mathbf{q}}|\lambda_{j,\mathbf{q}}|^2 }{jW} S_{j}^{z}$ which is $\sim |\lambda_{j}|^2 / (jW)$ appear. However, in the large tilt limit, this term negligibly contributes to the dominant term $M$.

Now that both $H_{XX}$ and $H_{XX}^{sb}$ don't appear in $D$, we next decouple the spin and phonon degrees of freedom using the following Polaron transformation, $a_{j,\mathbf{q}} = \tilde{a}_{j,\mathbf{q}} - \lambda_{j,\mathbf{q}}^{||} S_{j}^{z}$, and find $H_{ZZ}^{sb} + \sum_{j, \mathbf{q}} \hbar \omega_{j,\mathbf{q}}a_{j,\mathbf{q}}^{\dagger}a_{j,\mathbf{q}} = - \sum_{j} \frac{|\lambda_{j,\mathbf{q}}^{||}|^{2}}{\hbar \omega_{j,\mathbf{q}}} (S_{j}^{z})^{2} + \sum_{j, \mathbf{q}} \hbar \omega_{j,\mathbf{q}}\tilde{a}_{j,\mathbf{q}}^{\dagger}\tilde{a}_{j,\mathbf{q}} $. In other words, spin and phonon degrees of freedom are now decoupled with the spin-Hamiltonian $D$ exhibiting the form,
\begin{eqnarray}
 D = H_{ZZ} + \sum_{j} \left( \sum_{\mathbf{q}}|\lambda_{j,\mathbf{q}}^{||}|^{2}\right) (S_{j}^{z})^{2}.
\end{eqnarray}
Therefore, the effective spin Hamiltonian now becomes,
\begin{eqnarray}
    H_{\text{eff}}^{\prime} = \sum_{j} \big[ J S_{j}^{z}S_{j+1}^{z} - g_j (S_{j}^{z})^{2} + Wj S_{j}^{z}\big], 
\end{eqnarray}
where $g_{j} = \sum_{\mathbf{q}}\frac{|\lambda_{j,\mathbf{q}}^{||}|^{2}}{\hbar \omega_{j,\mathbf{q}}} $. $H_{\text{eff}}^{\prime}$ is a tilted Ising Hamiltonian with single ion anisotropy. To summarize, in the large tilt limit (above which the SMBL appears) the only remaining effect of spin-phonon interaction is the appearance of a single ion anisotropy in the effective Hamiltonian.

\end{document}